%% file: ICIP25_main.tex
\documentclass{article}
\usepackage{spconf}
\include{use_packages}
\include{my_def_preample}

\title{JOINT OPTIMIZATION OF PRIMARY AND SECONDARY TRANSFORMS USING RATE-DISTORTION OPTIMIZED TRANSFORM DESIGN} 
\name{Darukeesan Pakiyarajah$^{\star}$, Eduardo Pavez$^{\star}$, Antonio Ortega$^{\star}$, Debargha Mukherjee$^{\dagger}$, \\ \em{Onur Guleryuz$^{\dagger}$, Keng-Shih Lu$^{\dagger}$, Kruthika Koratti Sivakumar$^{\dagger}$} \begin{NoHyper}
\thanks{This work was supported in part by a grant from Google.}
\end{NoHyper}}
\address{$^{\star}$University of Southern California, Los Angeles, CA, USA \\
$^{\dagger}$Google LLC, Mountain View, CA, USA}

\begin{document}
\ninept
\maketitle
\begin{abstract}
Data-dependent transforms are increasingly being incorporated into next-generation video coding systems such as AVM, a codec under development by the Alliance for Open Media (AOM), and VVC. 
To circumvent the computational complexities associated with implementing non-separable data-dependent transforms, combinations of separable primary transforms and non-separable secondary transforms have been studied and integrated into video coding standards. 
These codecs often utilize rate-distortion optimized transforms (RDOT) to ensure that the new transforms complement existing transforms like the DCT and the ADST. 
In this work, we propose an optimization framework for jointly designing primary and secondary transforms from data through a rate-distortion optimized clustering. Primary transforms are assumed to follow a path-graph model, while secondary transforms are non-separable. We empirically evaluate our proposed approach using AVM residual data and demonstrate that 1) the joint clustering method achieves lower total RD cost in the RDOT design framework, and 2) jointly optimized separable path-graph transforms (SPGT) provide better coding efficiency compared to separable KLTs obtained from the same data.
\end{abstract}
\section{Introduction}
\label{sec:intro}
Using data-dependent transforms for image and video coding has been a topic of extensive research~\cite{Yeo12MDDT, fan2019signal, xu2012video, arrufat2014non}. Although the non-separable Karhunen-Loeve Transform (KLT) is theoretically optimal for image and residual blocks under various assumptions~\cite{Zhu11}, its practical adoption is limited due to memory constraints and higher computational complexity compared to trigonometric transforms, such as the DCT, which benefit from fast implementations. 
To address the limitations of KLT, several alternative approaches have been proposed, including separable KLTs~\cite{Yeo12MDDT, fan2019signal}, graph-based separable transforms (GBST)~\cite{Egilmez16gbst}, and secondary transforms\cite{Said16henst, Zhao18cpst, Zhao18jsnst, Zhao16nsst, Koo19, Saxena12stip, Zhao20stav1}. 
These methods aim to balance coding efficiency with computational complexity. Secondary transforms have been successfully integrated into VVC~\cite{zhao2021study}, while GBSTs and separable KLTs are being considered for next-generation codecs~\cite{zhao2021study, egilmez2022parametric}. 
Separable primary transforms, such as separable KLTs and GBSTs, are designed to be applied independently to the rows and columns of data blocks. On the other hand, secondary transforms are generally non-separable and are applied to a smaller subset of low-frequency primary transform coefficients to refine the coding efficiency. 
These systems typically use a rate-distortion optimized transform (RDOT) design approach~\cite{Zou13rdotLloyd, Zhao12rdot, Arrufat14rdot2}, which aims to create data-dependent transforms that minimize rate-distortion costs by using a clustering step to identify training data examples that cannot be well represented by the fixed transforms in the codec (e.g., DCT and ADST).

\begin{figure}[t]
    \centering
    \subfloat[\label{fig:indep_design}]{%
   \includegraphics[width = 0.24 \textwidth, trim={0.85cm 0.75cm 1.1cm 0.65cm}, clip ]{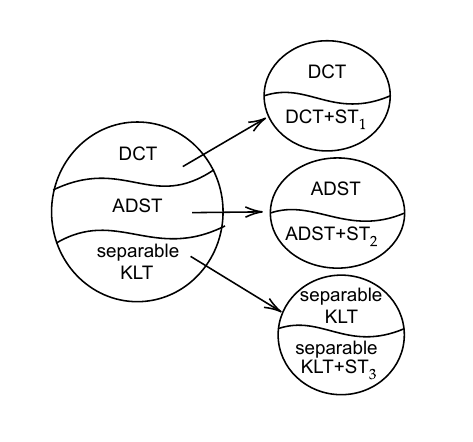}}
    % \hfill 
    \hspace{1cm}
  \subfloat[\label{fig:joint_design}]{%
        \includegraphics[width = 0.15 \textwidth, trim={0.2cm 2.7cm 0.8cm 0.0cm}, clip ]{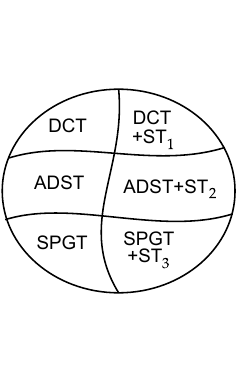} \vspace{4cm}
        }
    \vskip -2ex
    \caption{Example of clustering data in an RDOT design with two three primary transforms and three secondary transforms (ST): (a) Tree-structured clustering and (b) Joint clustering with separable path graph transform (SPGT).}        
    \vskip -2.5ex
    \label{fig:rdot_design}
\end{figure}

% This work addresses the problem of designing data-dependent primary (PT) and secondary transforms (ST) for intra-prediction residuals. Previous works, such as \cite{Ye08, Zou13rdotLloyd, Yeo12MDDT}, have focused on the design of primary transforms, while others, such as \cite{Said16henst, Zhao18cpst, Zhao18jsnst, Zhao16nsst, Koo19, Saxena12stip, Zhao20stav1}, have focused on designing secondary transforms. 

Current mode-dependent RDOTs for intra-prediction residuals have two major limitations. First, they are based on \textit{independently designed primary and secondary transforms}. 
As shown in \cref{fig:indep_design}, 
given the data, RDOT design and clustering are used to obtain a separable KLT and three clusters 
(corresponding to blocks using DCT, ADST, and the separable KLT). 
Subsequently, three independent RDOT design steps are performed for each cluster to design secondary transforms (ST$_1$, ST$_2$, and ST$_3$). 
%This results in the training data being divided into six clusters. 
In this tree-structured clustering, the primary transform is fixed, and the corresponding secondary transform design restricts the data in each cluster to either use a secondary transform or not, leading to a suboptimal design where the set of all primary and secondary transforms in the system may not effectively complement each other. 
A second limitation is that these RDOT design frameworks typically use \textit{separable KLTs as primary transforms}, requiring the learning of $N^2/2$ parameters from the data to construct an $N \times N$ transform matrix~\cite{Yeo12MDDT, Zhao12rdot}. 
Data can be scarce for learning a specific mode-dependent transform because some intra-prediction modes are infrequently used and, if multiple transforms are learned, available residual data needs to be further split into clusters \cite{Yeo12MDDT, Zou13rdotLloyd}. 
Consequently, the performance of learned separable KLTs may be limited due to insufficient training data. 

We address these limitations by (1) proposing a \textit{joint rate-distortion optimized clustering approach to learn primary and secondary transforms} and (2) employing learned \textit{separable path graph transforms (SPGT)} as primary transforms. We are unaware of other work considering joint primary and secondary transform design and leveraging path-graph models for mode-dependent RDOT design. 

In our joint design framework, all the data is clustered directly into six groups, allowing each data block to select the best combination of primary and secondary transforms from all choices in the system (see \cref{fig:joint_design}). 
The rate-distortion-optimized primary and secondary transforms are designed simultaneously, producing clusters different from existing tree-structured clustering systems, resulting in distinct data-dependent transforms. 
This unified framework ensures that combinations of primary and secondary transforms are optimized to complement each other, improving data representation efficiency. 
Furthermore, graph-based transform learning methods have recently been shown to have advantages over traditional KLT-based methods due to their robustness and ability to learn transforms from limited data~\cite{lu2024online}. We refer readers to the works in ~\cite{Egilmez16gbst, pavez2017learning} for a comprehensive review of graph-based transform designs applied to image and video coding. 
Moreover, for path graphs, direct learning of transform parameters from data can be performed without requiring covariance matrix estimation~\cite{lu2017closed}. 
Additionally, the path-graph model choice reduces the number of parameters learned from data to 
$N$ to design transforms for rows or columns with $N$ pixels, compared to $N^2/2$ in KLT-based methods. We leverage this in our joint learning framework to learn the SPGTs as primary transforms and address data scarcity challenges.

In designing secondary transforms, we aim to create data-dependent transforms for transform coefficients, which differ from the residual data used in primary transform design. 
The data scarcity problem remains a challenge and finding a suitable prior model for transform domain data is not as straightforward as in the case of rows/columns from residual data. To address the data scarcity problem, we follow previous works~\cite{Said16henst, Zhao18cpst, Zhao18jsnst, Zhao16nsst, Koo19, Saxena12stip, Zhao20stav1},  which select as a secondary transform a non-separable KLT that applies to the first $n\ll N^2$ low-frequency coefficients produced by the primary transform.  Furthermore, we observed that the covariance matrices of the primary transform coefficients are sparse, owing to the decorrelating effect of the primary transform. This sparsity further reduces the number of parameters and allows the corresponding KLTs to be learned from a smaller amount of data compared to the primary transforms. Thus, while non-separable KLTs would not be practical as primary transforms ($2048$ to be learned for $8 \times 8$ blocks), we use a non-separable KLT as a secondary transform with $n$ small enough to make learning the parameters possible with limited amounts of data. For example, if $ n=16$, fewer than $128$ parameters need to be learned due to the sparse structure of the covariance matrix. To the best of our knowledge, non-separable KLTs are always used in practice as the secondary transform~\cite{bross2021overview,zhao2021study}.
%\footnote{For intra-prediction mode D\_67, one of the less frequent modes in AVM, we get a cluster with $3964$ samples of $8\times8$ blocks, non-separable KLT for all $64$ coefficients require learning $2048$ parameters whereas a secondary transform for only $16$ low-frequencies requires learning $128$ parameters. According to a general $10$ times rule on sufficient data per parameter, this training sample is insufficient for the first and sufficient for the second.}. 
%Following this approach, we use non-separable KLTs to design secondary transforms in our joint design framework.

We empirically evaluate our proposed design of mode-dependent primary and secondary transforms using intra-prediction residual data from AVM, a codec under development by the Alliance for Open Media (AOM). Our results demonstrate that the joint clustering approach with SPGTs achieves the lowest RD cost in transform design, outperforming existing tree-structured clustering methods with separable KLTs and providing an average bitrate saving of $0.3\%$ compared to tree-structured clustering, all without introducing additional implementation complexity. Furthermore, SPGTs achieve bitrate savings ranging from $0.5\%$ to $3.83\%$ over KLT-based primary transform designs, highlighting their effectiveness in learning primary transforms from limited data.

\section{Rate-Distortion Optimized Transforms}
Assume we have $M$ residual blocks of dimension $N \times N$, $\{\Xm_i\}_{i=1}^M$, and $K$ transforms, $\{\Fm_j\}_{j=1}^K$. Let $\tilde{\Xm}_{i,j}$ denote the transform coefficients after applying the transform $\Fm_j$ and $d(\Xm_i, \Fm_j)$ represent the distortion due to entrywise quantization of $\tilde{\Xm}_{i,j}$. 
Furthermore, let $B_{i,j}$ denote the number of bits required to encode  $\tilde{\Xm}_{i,j}$, including both overhead and coefficient bits. 
In the rate-distortion (RD) optimization step of the residual coding process, the transform $\Fm_j$ that minimizes an RD  cost is selected for each block $\Xm_i$. 
Let $\Sc_j$ denote the index set of residual blocks that choose the transform $\Fm_j$: 
\begin{equation} 
\Sc_j = \left\{ i : \text{arg }\underset{k}{\text{min }} d(\Xm_i, \Fm_k) + \lambda B_{i,k} = j \right\}, \label{eq:indexset} 
\end{equation} 
where $\lambda\geq 0$ is the predefined Lagrange multiplier commonly used in video codecs to control 
the RD trade-off~\cite{ringis2023disparity}. 
The RDOT design problem aims to perform clustering and transform design simultaneously, which can be formulated as~\cite{Zou13rdotLloyd, Arrufat14rdot2, Zhao12rdot}
\begin{align}
    \label{eq:rdotprb}
      \underset{\{\Fm_j\}_{j=1}^K}{\text{min }} & \sum_{j=1}^K  \sum_{i\in\Sc_j} d(\Xm_i,\Fm_j) + \lambda B_{i,j} \\
      \text{s.t.}\:\: & \Fm_j \in \Cc_j \quad \text{for } j = 1, \hdots, K, \notag
\end{align}
where $\Cc_j$ denotes the set of transforms; for instance, $\Cc_j$ could be the set of separable transforms. 

\cref{alg:rdot_algo}, a Lloyd-type algorithm that has been employed in~\cite{Zou13rdotLloyd, Arrufat14rdot2, Zhao12rdot}, solves \eqref{eq:rdotprb} by iteratively alternating between two steps: (1) updating the transforms given the clusters based on the cluster-specific update rules and (2) updating the clusters given the transforms, where each residual block is reassigned to the cluster (transform) that minimizes its contribution to the overall objective. These steps are repeated until there is no significant change in the objective cost, indicating convergence. 
This approach has been used to design  primary transforms based on  separable KLTs   \cite{Arrufat14rdot2} 
and secondary non-separable KLTs \cite{Said16henst, Zhao18jsnst}. 
In this work, we use this algorithm to simultaneously design SPGTs as primary transforms and non-separable KLTs as secondary transforms.

\begin{algorithm}[t]
\begin{algorithmic}[1]
% \label{algorithm:Algo1}
\caption{RDOT Learning} \label{alg:rdot_algo}
\renewcommand{\algorithmicrequire}{\textbf{Input:}}
 \renewcommand{\algorithmicensure}{\textbf{Output:}}
 \REQUIRE $\{\Xm_i\}_{i=1}^M$, $\Cc_j$ for $j=1, \hdots, K$
 \renewcommand{\algorithmicrequire}{\textbf{Initialize:}}
 \REQUIRE $\{\Sc_j\}_{j=1}^K$, $\{\Fm_j\}_{j=1}^K$
            \REPEAT
                \STATE update $\Fm_j \in \Cc_j$  for $j=1, \hdots, K$ \label{alg_step:tup}
                \STATE $\Sc_j = \left\{ i : \text{arg }\underset{k}{\text{min }} d(\Xm_i,\Fm_k) + \lambda B_{i,k} = j \right\}$ \\ \qquad for $j=1, \hdots, K$ \label{alg_step:cluster}
            \UNTIL{convergence}
\ENSURE  $\{\Fm_j\}_{j=1}^K$
\end{algorithmic}
\end{algorithm}

\section{Joint Optimization of Primary and Secondary Transforms}
\label{sec:Jointrdot}
%In this work, we propose 1) a joint-clustering approach for residual data, considering all primary and secondary transforms simultaneously using ~\cref{alg:rdot_algo}, and 2) designing separable primary transforms under the assumption that the residual data follows graph-based models, allowing for more efficient transform learning. 
% In this section, we describe the transform update rules and clustering method used in our joint design framework for simultaneously learning SPGTs and secondary transforms.
% \subsection{Transform update rules}

For a block $\Xm_i$, a separable transform can be defined using two orthonormal $N \times N$ transforms,  a column transform $\Cm_j$ and a row transform $\Rm_j$. The transform domain representation of $\Xm_i$ is given by $\hat\Xm_{i,j} = \Cm_j\tr\Xm_i\Rm_j$. Let $\xv_{i} = \vec(\Xm_{i})$ represent the vector formed by stacking the columns of the matrix $\Xm_{i}$. Then, the transform domain vector can be expressed as:
\begin{equation} 
    \hat\xv_{i,j} = \vec(\hat\Xm_{i,j}) = \Gm_j\tr\xv_i, 
\end{equation} 
where $\Gm_j = \Rm_j \otimes \Cm_j$ denotes the separable primary transform, and $\otimes$ denotes the Kronecker product operation on two matrices. Let $\Pm_j$ represent the $N^2 \times N^2$ permutation matrix that reorders the elements of $\hat\xv_{i,j}$ according to a scanning order (e.g., zig-zag scanning order). 
Suppose a non-separable orthonormal secondary transform, denoted by an $n \times n$ matrix $\tilde\Tm_j$, is applied to the first $n$ low-frequency coefficients after reordering. 
Then, the transform coefficients obtained by applying the separable primary transform followed by the non-separable secondary transform are given by $\hat\yv_{i,j} = \Tm_j\tr\Pm_j\tr\Gm_j\tr\xv_i$, where $\Tm_j \in \Tc$ is an $N^2 \times N^2$ block matrix, and  
\begin{equation}
    \Tc = \left\{ \Tm :  \Tm = \begin{bmatrix} \tilde\Tm & \bf 0 \\ \bf 0 & \Id \end{bmatrix}, \: \tilde\Tm\tr\tilde\Tm =\Id \right\}.
\end{equation}
The inverse transform is $\xv = \Fm_j\hat\yv_{i,j}$, where $\Fm_j=\Gm_j\Pm_j\Tm_j$.

After entrywise quantization of $\yv_{i,j}$, we get $\hat \yv_{i,j}^Q$. 
Then, the distortion of the block due to quantization is expressed as $d(\Xm_i,\Fm_j) = \|\xv_i - \Fm_j \hat \yv_{i,j}^Q\|_2^2$. During the entropy coding process with coders such as CABAC~\cite{Marpe01cabac}, $B_{i,j}$ is highly context-dependent, making it difficult to determine an accurate proxy for $B_{i,j}$ and adding complexity to the RD optimized clustering problem. To address this challenge, in \cite{Zhao12rdot, Zou13rdotLloyd}, the authors proposed using the $\ell_0$-norm of the quantized coefficients, $\|\hat\yv_{i,j}^Q\|_0$, which represents the number of non-zero coefficients, as a proxy for $B_{i,j}$ in the RDOT framework. We use the same proxy in our joint learning framework, so the RD optimized clustering in step~\ref{alg_step:cluster} of ~\cref{alg:rdot_algo} becomes: 
\begin{equation}
    \Sc_j = \left\{ i : \text{arg }\underset{k}{\text{min }} \|\xv_i - \Fm_k \hat\yv_{i,k}^Q\|_2^2 + \lambda \|\hat\yv_{i,k}^Q\|_0 = j \right\}.
\end{equation} 
%Consequently, the transform learning problem for step~\ref{alg_step:tup} of \cref{alg:rdot_algo} is formulated as:
%\begin{equation} 
%\Fm_j = \text{arg }\underset{\Fm \in \Cc_j}{\text{min }} %\sum_{i \in \Sc_j} \|\xv_i - \Fm \hat \yv_{i,j}^Q \|_2^2, 
%\label{eq:tr_learn}
%\end{equation}
%which is similar to the problem of separable transform learning problem studied in the literature~\cite{Zou13rdotLloyd, Yeo12MDDT, Egilmez16gbst}. 

In our joint learning framework, we define six sets of transforms that are used for the transform updates in step \ref{alg_step:tup} of \cref{alg:rdot_algo}:
\begin{align}
    \Cc_1 & = \left\{ \Fm :  \Fm = \Gm_\text{DCT}, \Gm_\text{DCT} = \Rm_\text{DCT} \otimes \Cm_\text{DCT} \right\}, \notag \\
    \Cc_2 & = \left\{ \Fm :  \Fm = \Gm_\text{ADST}, \Gm_\text{ADST} = \Rm_\text{ADST} \otimes \Cm_\text{ADST} \right\}, \notag \\
    \Cc_3 & = \left\{ \Fm :  \Fm = \Gm_\text{SPGT} = \Rm_3 \otimes \Cm_3, \Rm_3\tr\Rm_3=\Cm_3\tr\Cm_3 = \Id \right\} \cap \Gc, \notag \\
    \Cc_4 & = \left\{  \Fm :  \Fm = \Fm_1\Pm_4\Tm_4,  \Pm_4 \text{ fixed}, \Tm_4 \in \Tc \right\}, \notag \\
    \Cc_5 & = \left\{  \Fm :  \Fm = \Fm_2\Pm_5\Tm_5, \Pm_5 \text{ fixed}, \Tm_5 \in \Tc \right\},  \notag \\
    \Cc_6 & = \left\{  \Fm :  \Fm = \Fm_3\Pm_6\Tm_6,  \Pm_6 \text{  fixed}, \Tm_6 \in \Tc \right\},\notag
\end{align}
 where $\Gc$ denotes the path graph model constraints. Thus, the primary transforms can be either DCT, ADST or a learned SPGT. The sets $\Cc_1$ and $\Cc_2$ form a primary/secondary pair, and the same for the sets $\Cc_2/\Cc_5$ and $\Cc_3/\Cc_6$. Detailed descriptions of how SPGTs and secondary transforms are learned can be found in \cref{sec:gbst_sol} and \cref{sec:sec_sol}, respectively.

\subsection{Separable path graph transform}
\label{sec:gbst_sol}
\begin{figure}
    \centering
    \includegraphics[width=1\linewidth,  trim={0cm 0.4cm 0cm 1cm}, clip ]{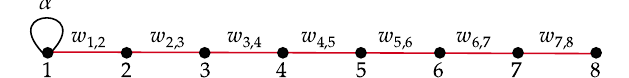}
    \vskip -1ex
    \caption{A path graph model with $N=8$ nodes, self-loop with weight $\alpha$ and edge weights $w_{i,j}$.}
    \vskip -2.5ex
    \label{fig:path_graph}
\end{figure}
Let us consider the set of SPGTs, $\Cc_3$, and the corresponding data cluster $\Sc_3$. In this case, we learn the inverse covariances of the column and row data, $\Lm_\text{col}$ and $\Lm_\text{row}$, respectively, instead of the covariance matrices as in separable KLT methods. We adopt a model based on a path graph with a self-loop at the first node for the inverse covariance matrices of the rows and columns of  $\{ \Xm_i \}_{i \in \Sc_3}$.  
This path graph model is parameterized by a self-loop weight $\alpha$ and edge weights $w_{i,j}$, as illustrated in \cref{fig:path_graph}. The choice of this path graph model is motivated by the fact that a path graph with no self-loop and all edge weights equal to $1$ corresponds to DCT, while adding a self-loop with weight $1$ at the first node corresponds to ADST. In \cite{lu2017closed}, it has been shown that the solution of the graph learning problem in \cite{Egilmez16gbst} is provided by closed-form formulas giving the edge weights and the self-loop weight:
\begin{align} 
    w_{i,j} & = \left[\frac{1}{P}\sum_{p=1}^P (x_p(i)-x_p(j))^2 + \beta \right]^{-1}, \notag \\
    \alpha & = \left[\frac{1}{P}\sum_{p=1}^P x_p^2(1)\right]^{-1}, \notag
\end{align}
where $P$ is the number of training samples used to learn the model parameters, $x_p(i)$ represents the data observed at the $i^\text{th}$ node of the graph for the $p^\text{th}$ training sample, and $\beta>0$ is a small positive number to avoid infinity. Then the transforms $\Cm_3$ and $\Rm_3$ are given by the eigenbasis of $\Lm_\text{col}$ and $\Lm_\text{row}$ respectively, and the SPGT corresponding to this cluster is given by $\Gm_\text{SPGT} = \Rm_3 \otimes \Cm_3$.

Note that in this case, we obtain an $N \times N$ transform by learning only $N$ parameters from the data, $\alpha$ and $w_{i,j}$, whereas the separable KLT approaches in \cite{Zhao12rdot,Yeo12MDDT} need to learn $N^2/2$ parameters. This significant reduction in the number of learned parameters can be particularly beneficial in data-scarce scenarios, such as when dealing with less frequent intra-prediction modes or smaller clusters in the clustering framework, where sufficient training data is limited.
\subsection{Secondary transform}
\label{sec:sec_sol}
Consider the case where the transform corresponding to the $j^\text{th}$ cluster is a primary transform followed by a secondary transform, with $\Gm_j$ and $\Pm_j$ fixed a priori. This set of transforms can be expressed by
\begin{equation}
    \Cc_j = \left\{ \Fm :  \Fm = \Gm_j\Pm_j\Tm_j, \Gm_j, \Pm_j \text{ are fixed}, \Tm_j \in \Tc\right\}.
    \label{eq:cons_sec}
\end{equation}
Here, $\Gm_j$ could be any predefined separable transform. We use an empirical approach to obtain $\Pm_j$ for each intra-prediction mode in our setting. We apply the fixed separable primary transform, $\Gm_j$, assigned to this cluster to $\{\Xm_i\}_{i \in \Sc_j}$. Then, we compute the sample variance of each of these transform coefficients. The sorting order of the transform coefficients based on their decreasing sample variance determines the permutation matrix $\Pm_j$. Then, the secondary transform is obtained by solving the following optimization \cite{Zhao16nsst, Zhao18cpst, Zhao18jsnst}: 
\begin{align}
    \Fm_j &=\text{arg }\underset{\Fm \in \Cc_j}{\text{min }} \sum_{i \in \Sc_j} \|\xv_i - \Fm \hat \yv_{i,j}^Q \|_2^2 \notag \\
    &= \Gm_j\Pm_j\left[\text{ arg }\underset{\Tm_j \in \Tc}{\text{min }} \sum_{i \in \Sc_j} \| \hat\zv_{i,j} - \Tm_j \hat\zv_{i,j}^Q\|_2^2\right],
    \label{eq:sec_prb}
\end{align}
where $\hat\zv_{i,j} = \Pm_j\tr\Gm_j\tr\xv_i$, the second equality follows from the Parseval's identity. In ~\cite{Zhao16nsst, Zhao18cpst, Zhao18jsnst}, it has been shown that the solution $\Tm_j$ in the above problem is obtained by first computing the non-separable secondary transform $\tilde\Tm_j$ as the non-separable KLT of the $n$ significant frequency coefficients derived from all the residual data in this cluster and then, $\Tm_j$ is constructed from $\tilde\Tm_j$ such that $\Tm_j \in \Tc$.

\section{Empirical results}
\label{sec:empRes}
We use intra-prediction residuals obtained from AVM for both learning and testing transforms in our experiments. 
\subsection{Mode-dependent RDOT learning}
\begin{figure} 
    \vskip -2ex
    \centering
    \subfloat[\label{fig:rd_total_m2}]{%
   \includegraphics[width = 0.24 \textwidth, trim={0.1cm 0.0cm 0.8cm 0.20cm}, clip ]{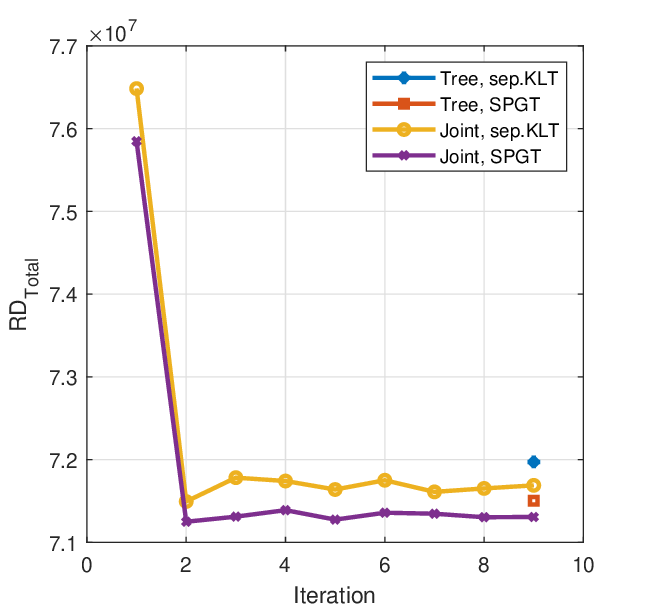}}
    \hfill
  \subfloat[\label{fig:rd_total_m4}]{%
   \includegraphics[width = 0.24 \textwidth, trim={0.1cm 0.0cm 0.8cm 0.20cm}, clip ]{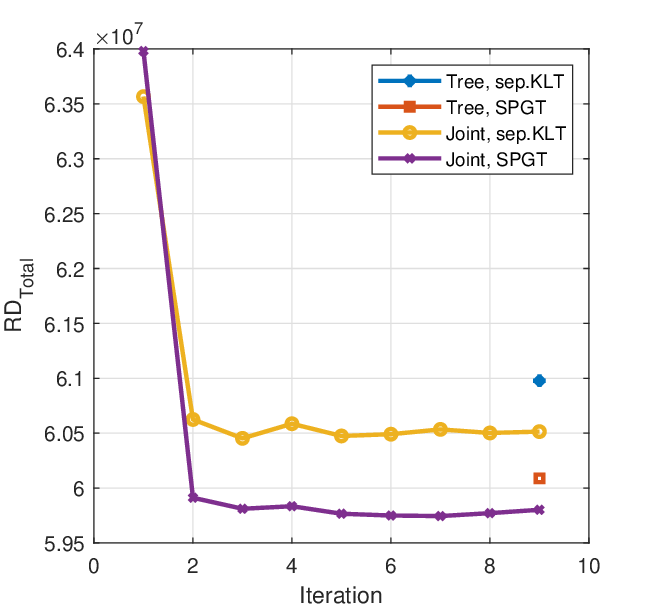}}
    \vskip -2.5ex
    \caption{Total RD cost achieved by the four RDOT learning methods for (a) H\_PRED and (b) D\_135\_PRED. For tree-structured clustering, $\text{RD}_\text{Total}$ is computed after designing all transforms and represented as points.}        
    \vskip -3.5ex
    \label{fig:rd_total}
\end{figure}

We design data-dependent primary and secondary transforms for the 12 principal intra-prediction modes in AVM in a mode-dependent manner. In this setup, each intra-prediction mode is allowed to use six $(K=6)$ transforms, and the joint RDOT learning is performed as described in \cref{sec:Jointrdot}. We refer to our approach as joint clustering with SPGT (Joint, SPGT).

\begin{figure} 
    \centering
    \subfloat[\label{fig:var_order_mode1}]{%
   \includegraphics[width = 0.24 \textwidth, trim={1.3cm 1.2cm 1.2cm 0.8cm}, clip ]{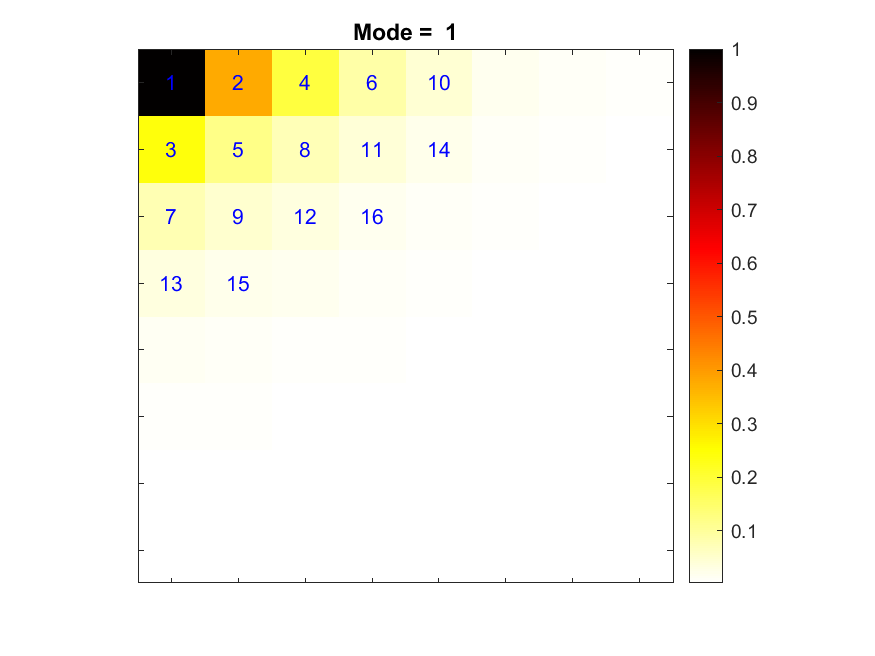}}
    \hfill
  \subfloat[\label{fig:var_order_mode2}]{%
        \includegraphics[width = 0.24 \textwidth, trim={1.3cm 1.2cm 1.2cm 0.8cm}, clip ]{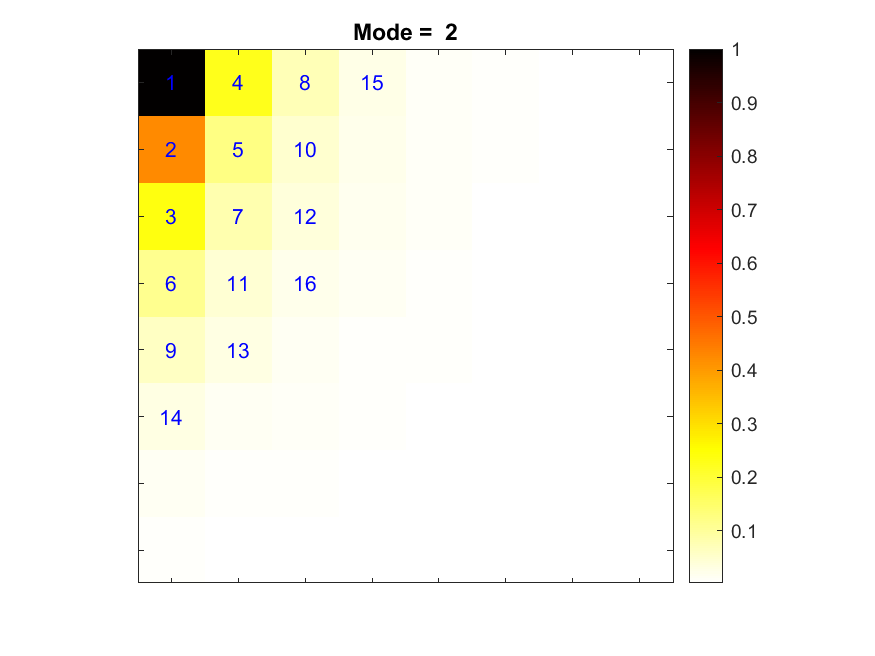}}
    \vskip -2ex
    \caption{Sample variance (normalized) of the transform coefficients after applying separable DCT, corresponding to intra-prediction modes: (a) V\_PRED), and (b) H\_PRED. The numbers $1$–$16$ indicate the coefficients selected for the secondary transform.}        
    \vskip -3.5ex
    \label{fig:scan_order}
\end{figure}

First, we show how the low-frequency coefficients are selected for different intra-prediction modes for the subsequent secondary transform.  Examples of the sample variance of the transform coefficients obtained for two different intra-prediction modes of AVM after applying $\Gm_\text{DCT}$ as the primary transform are shown in \cref{fig:scan_order}. This highlights the importance of adopting a mode-dependent scanning order to select the low-frequency coefficients for the subsequent secondary transform rather than relying on a fixed order for all modes. Examples of the $16$ coefficients selected for two different $8 \times 8$ intra-prediction modes of AVM are illustrated in \cref{fig:scan_order}. 

\begin{figure}
    \centering
    \includegraphics[width=0.9\linewidth,  trim={1.2cm 1.2cm 0cm 0.6cm}, clip ]{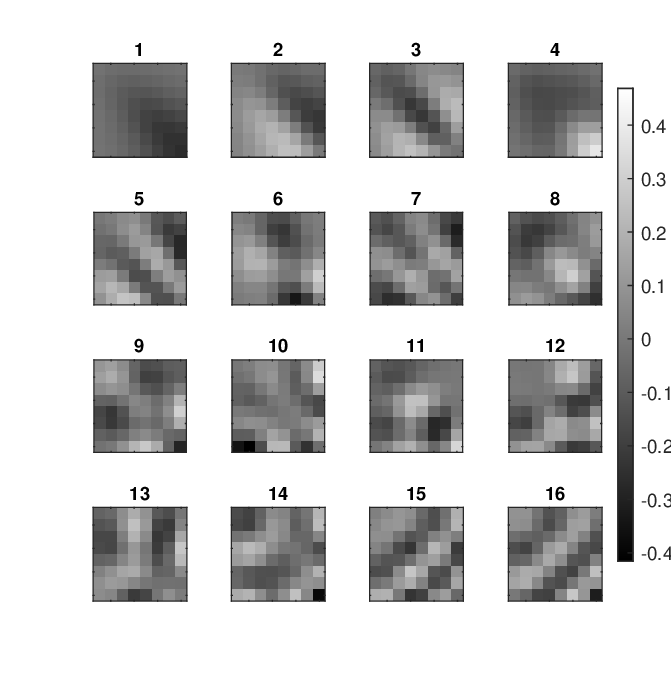}
    \vskip -1ex
    \caption{First $16$ columns (basis) of the transform $\Fm_6 = \Gm_\text{SPGT}\Pm_6\Tm_6$ designed using the joint learning with SPGT for D\_135\_PRED (diagonal prediction), $8 \times 8$ residuals, reshaped into $8 \times 8$ blocks.}
    \vskip -4.5ex
    \label{fig:basis}
\end{figure}

Moreover, to compare against the existing tree-structured clustering approaches~\cite{Zhao16nsst, Zhao20stav1}, we design primary and secondary transforms following these steps:
\begin{enumerate}[topsep=-1pt, itemsep=1ex, partopsep=-1ex, parsep=-1ex]
    \item Run \cref{alg:rdot_algo} with $\{\Xm_i\}_{i=1}^M$, $\Cc_1$, $\Cc_2$, and $\Cc_3$ as inputs to obtain $\Gm_\text{SPGT}$ and the clusters $\Sc_1$, $\Sc_2$, and $\Sc_3$.
    \item Run three independent instances of \cref{alg:rdot_algo} to design the secondary transforms with (i) $\{\Xm_i\}_{i\in\Sc_1}, \Cc_1, \Cc_4$, (ii) $\{\Xm_i\}_{i\in\Sc_2}, \Cc_2, \Cc_5$, (iii) $\{\Xm_i\}_{i\in\Sc_3}, \Cc_3, \Cc_6$ as inputs to obtain secondary transforms.
\end{enumerate}
We refer to this method as tree-structured clustering with SPGT (Tree, SPGT). To compare the SPGTs against the separable KLTs method for designing primary transforms, we repeat the joint and tree-structured clustering of the same data, but with $\Fm_3$ obtained using the separable KLT method as described in ~\cite{Yeo12MDDT}, where the $\Rm_3$ and $\Cm_3$ are obtained from the eigenbasis of $\Sm_\text{col}$ and $\Sm_\text{row}$, the sample covarianve matrices of rows and columns of $\{\Xm_i\}_{i\in\Sc_3}$, respectively. 

Note that we have two clustering approaches, $\{\text{Joint}, \text{Tree}\}$, and two transform learning methods to design primary transforms, $\{\text{SPGT}, \text{sep.KLT}\}$, with each combination of clustering and transform learning resulting in four methods to design RDOTs: (Joint, SPGT), (Joint, sep.KLT), (Tree, SPGT), and (Tree, sep.KLT). For our experiments, we consider training data consisting of $8 \times 8$ and $16 \times 16$ residuals and train mode-dependent transforms separately for each block size. The number of coefficients selected for secondary transforms is set to $n=16$ for $8 \times 8$ blocks and $n=64$ for $16 \times 16$ blocks. Furthermore, in RDOT design, we use $QP=28$ and $\lambda = 0.85\times 2^{(QP-12)/3}$, which are typical values used for RDOT design for intra-prediction residuals in codecs like AVC and HEVC~\cite{Zhao12rdot, Arrufat14rdot2}. In our experiments with different values of QP (or $\lambda$), we observed that training the transforms with a smaller QP (smaller $\lambda$) from the typical range of QP values used in compression applications allows the same transforms to be used across different QP values. This eliminates the need to train transforms for different QP values without a significant trade-off in bitrate savings. Furthermore, for each cluster corresponding to the secondary transforms, we observed that the covariance matrices of the $n$ low-frequency transform coefficients are of full rank, substantiating that non-separable KLTs with fewer parameters can effectively learn the secondary transforms. We omit the detailed experimental results due to space limitations.

The first $16$ basis functions of the transform $\Fm_6 = \Gm_\text{SPGT}\Pm_6\Tm_6$ designed using the joint learning method for D\_135\_PRED (diagonal prediction), $8 \times 8$ residuals are shown in \cref{fig:basis}. From the figure, we observe that the low-frequency basis functions exhibit smooth variations in the direction of the corresponding intra-prediction mode, demonstrating that the combination of primary and secondary transforms effectively adapts to the statistics of the data. Further, to compare these four learning methods, we compute the total RD cost obtained using these methods, given by
\begin{equation}
    \text{RD}_\text{Total} = \sum_{j=1}^K  \sum_{i\in\Sc_j} \|\xv_i - \Fm_j \hat\yv_{i,j}^Q\|_2^2 + \lambda \|\hat\yv_{i,j}^Q\|_0, 
\end{equation}
for $8 \times 8$ intra-prediction residuals corresponding to all the intra-prediction modes. The results for two of those modes are presented in \cref{fig:rd_total}. From this figure, we observe the following: 1) both joint clustering methods achieve a lower total RD cost compared to their respective tree-structured clustering methods, and 2) methods employing SPGTs achieve a lower objective cost compared to those using separable KLTs. These observations are consistent across all intra-prediction modes and provide empirical evidence supporting our claims regarding the benefits of joint clustering and the advantages of path graph-based transforms.

\vskip -2ex
\subsection{Testing}
% \begin{figure} 
%     \vskip -2ex
%     \centering
%     {%
%    \includegraphics[width = 0.4 \textwidth, trim={0.75cm 0.1cm 1.0cm 0.4cm}, clip ]{Figures/data_bd8.eps}}
%     \hfill
%       \vskip -2ex
%     \caption{BD-rate saving vs the number of data samples used to train RDOTs for (\%) $8 \times 8$ residual blocks. The points along the same vertical line correspond to the same (labeled) intra-prediction mode, with more negative values indicating greater bitrate savings.}        
%     \vskip -1.5ex
%     \label{fig:dz_bd8}
% \end{figure}

% \begin{figure}
%     \centering
%   {%
%    \includegraphics[width = 0.4 \textwidth, trim={0.75cm 0.1cm 1.0cm 0.4cm}, clip ]{Figures/data_bd16.eps}} \hfill
%     \vskip -2ex
%     \caption{BD-rate saving vs the number of data samples used to train RDOTs for $16 \times 16$ residual blocks.}        
%     \vskip -3.5ex
%     \label{fig:dz_bd}
% \end{figure}

\begin{figure} 
    \vskip -2ex
    \centering
    \subfloat[\label{fig:dz_bd8x8}]{%
   \includegraphics[width = 0.24 \textwidth, trim={0.75cm 0.1cm 1.2cm 0.4cm}, clip ]{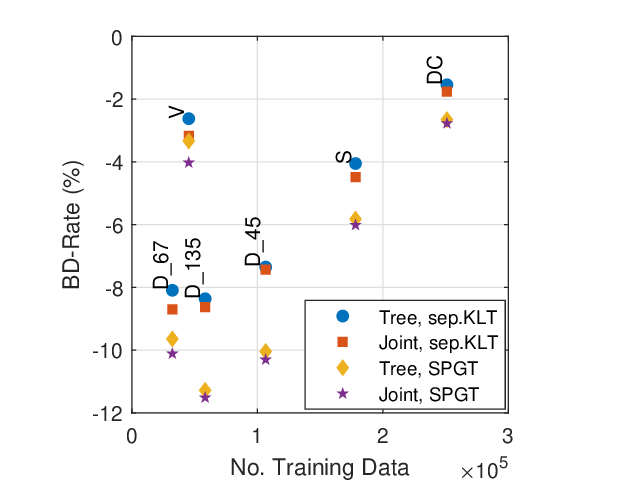}}
    \hfill
  \subfloat[\label{fig:dz_bd16x16}]{%
   \includegraphics[width = 0.24 \textwidth, trim={0.75cm 0.1cm 1.2cm 0.4cm}, clip ]{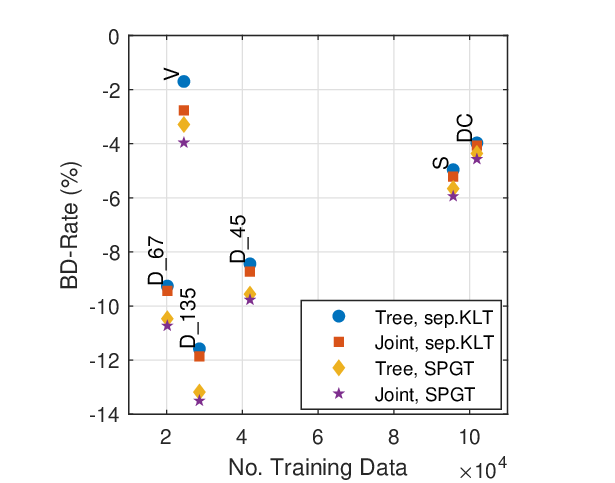}}
    \vskip -2ex
    \caption{BD-rate saving vs the number of data samples used to train RDOTs for (\%) (a) $8 \times 8$ residual blocks and (b) $16 \times 16$ residual blocks. The points along the same vertical line correspond to the same (labeled) intra-prediction mode, with more negative values indicating greater bitrate savings.}        
    \vskip -3.5ex
    \label{fig:dz_bd}
\end{figure}

\begin{table*}[t]
  \centering
  \resizebox{\textwidth}{!}{%
  \begin{tabular}{|l *{13}{|c}|}
    \hline
    Method &DC & V & H & D$\_45$ & D$\_135$ & D$\_113$ & D$\_157$ & D$\_203$ & D$\_67$ & S & S$\_$V & S$\_$H \\ %& Average\\
    \hline
    Tree, sep. KLT  & $-1.54$ & $-2.62$ & $-2.43$ & $-7.35$ & $-8.36$ & $-6.48$ & $-5.73$ & $-5.94$ & $-8.09$ & $-4.05$ & $-3.78$ & $-4.63$ \\ % & $-5.11 \pm 2.21$ \\ 
    \hline
    Joint, sep. KLT & $-1.76$ & $-3.17$ & $-2.78$ & $-7.43$ & $-8.63$ & $-6.90$ & $-5.90$ & $-6.64$ & $-8.70$ & $-4.48$ & $-3.90$ & $-4.83$ \\ %& $-5.40 \pm 2.17$ \\
    \hline
    Tree, SPGT & $-2.65$ & $-3.33$ & $-4.97$ & $-10.04$ & $-11.28$ & $-9.14$ & $-9.15$ & $-8.92$ & $-9.64$ & $-5.82$ & $-5.09$ & $-6.76$ \\ %& $-7.23 \pm 2.71$ \\
    \hline
    Joint, SPGT & $-2.77$ & $-4.02$ & $-5.32$ & $-10.30$ & $-11.51$ & $-9.32$ & $-9.73$ & $-9.48$ & $-10.11$ & $-6.01$ & $-5.12$ & $-7.00$ \\ %& $-7.56 \pm 2.74$ \\
    \hline
  \end{tabular}
  }
  \vskip -1.5ex
  \caption{BD rate savings (\%) with respect to the baseline %for four methods, : independent learning with separable KLT (Indep. sep. KLT), joint learning with separable KLT (Joint sep. KLT), independent learning with SPGT (Indep. SPGT), and joint learning with SPGT (Joint SPGT), 
  for $8 \times 8$ residuals. More negative values indicate greater bitrate savings.}
  \label{tab:8x8bdrate}
    \vskip -1ex
\end{table*}

\begin{table*}[t]
  \centering
    \resizebox{\textwidth}{!}{%
  \begin{tabular}{|l *{13}{|c}|}
    \hline
    Method &DC & V & H & D$\_45$ & D$\_135$ & D$\_113$ & D$\_157$ & D$\_203$ & D$\_67$ & S & S$\_$V & S$\_$H \\ %& Average\\
    \hline
    Tree, sep. KLT  & $-3.97$ & $-1.70$ & $-2.21$ & $-8.44$ & $-11.58$ & $-9.70$ & $-8.52$ & $-8.34$ & $-9.26$ & $-4.96$ & $-1.89$ & $-3.90$  \\ %& $-6.21 \pm 3.32$ \\ 
    \hline
    Joint, sep. KLT & $-4.07$ & $-2.77$ & $-2.91$ & $-8.73$ & $-11.86$ & $-10.10$ & $-8.79$ & $-8.57$ & $-9.44$ & $-5.22$ & $-3.52$ & $-4.41$ \\ %& $-6.70 \pm 3.05$ \\
    \hline
    Tree, SPGT & $-4.36$ & $-3.29$ & $-3.88$ & $-9.56$ & $-13.18$ & $-10.67$ & $-9.89$ & $-9.62$ & $-10.47$ & $-5.66$ & $-3.85$ & $-5.60$ \\ %& $-7.50 \pm 3.25$ \\
    \hline
    Joint, SPGT & $-4.57$ & $-3.96$ & $-4.16$ & $-9.77$ & $-13.50$ & $-10.85$ & $-9.98$ & $-9.73$ & $-10.73$ & $-5.94$ & $-4.78$ & $-5.93$ \\ %& $-7.82 \pm 3.13$ \\
    \hline
  \end{tabular}
  }
  \vskip -1.5ex
  \caption{BD rate savings (\%) with respect to the baseline 
  %for four methods: independent learning with separable KLT (Indep. sep. KLT), joint learning with separable KLT (Joint sep. KLT), independent learning with SPGT (Indep. SPGT), and joint learning with SPGT (Joint SPGT), 
  for $16 \times 16$ residuals. More negative values indicate greater bitrate savings.}
  \label{tab:16x16bdrate}
    \vskip -1.5ex
\end{table*}

Next, we demonstrate that the transforms designed using the joint clustering method with SPGT provide better coding gains than other methods in video coding applications. For this purpose, we use a different test dataset of residuals from AVM. For each intra-prediction mode, we independently encode each residual block. The process involves applying the transform, uniform quantization of transform coefficients with quantization parameters $QP=\{ 26,27,28,29,30,31 \}$, and entropy coding the coefficients using a CAVLC encoder~\cite{karczewicz2004context}. We first select the best primary transform from the three primary transforms for each block through a standard RDO process. Then, we perform a secondary RDO step to decide whether or not to apply a secondary transform. We use a fixed $3$-bit overhead to signal the selected transform for each block: two bits to indicate the primary transform and one bit to signal whether a secondary transform is applied. This RDO approach for choosing the best combination of primary and secondary transforms is consistent with state-of-the-art codecs using secondary transforms.

We compare all four learning methods against a baseline method that uses only two primary transforms, DCT and ADST, with RDO, quantization, and entropy coding performed as described. The baseline has only two primary transforms and uses only a one-bit overhead to signal the selected transform. For each method, including the baseline, we compute the bitrate and distortion for each block and average the results across multiple blocks. We present the average Bjontegaard rate (BD-rate) savings~\cite{bjontegaard2001calculation} compared to the baseline for each intra-prediction mode of $8 \times 8$ blocks in \cref{tab:8x8bdrate} and $16 \times 16$ blocks in \cref{tab:16x16bdrate}. Note that since all transform learning is performed offline, the bitrate reductions come at no additional implementation cost. Furthermore, most of the gain in the transform design is achieved by using SPGTs. Significant gains in directional prediction modes suggest that data-dependent transforms effectively adapt to their directional statistics. An average of $0.3\%$ bitrate reduction from the joint clustering compared to tree-structured clustering indicates that the primary and secondary transforms are better optimized to complement each other compared to tree-structured clustering. Furthermore, we present the BD-rate savings for all methods as a function of the number of data samples used to train the RDOTs in \cref{fig:dz_bd8x8} and \cref{fig:dz_bd16x16}. For modes with fewer training samples, the difference between KLT-based methods and SPGTs is significant, while the difference diminishes for modes with larger amounts of training data. This highlights the advantage of using path graph prior to learn transforms with fewer parameters in data-scarce scenarios.

%Additionally, we compute the average bitrate and PSNR for all $8 \times 8$ residual blocks and $16 \times 16$ residual blocks, and the corresponding RD curves are shown in \cref{fig:rd_curves} (a) and (b) respectively. From the tables, we can observe that graph-based learning provides significant bitrate saving compared to separable KLT-based methods for directional prediction modes, and joint clustering provides $0.3\%$ bitrate saving on average compared to tree-structured clustering methods in both $8 \times 8$ case $16 \times 16$ case. 

\vskip -2ex
\section{Conclusion}
In this work, we proposed a method for simultaneously learning data-dependent primary and secondary transforms using the rate-distortion optimized transform design framework. To learn the separable primary transforms, we utilized a path graph model with a self-loop at the first node, enabling efficient transform learning with fewer parameters in data-scarce scenarios. We empirically evaluated our approach using AVM intra-prediction residuals, comparing joint versus tree-structured clustering and separable KLTs versus SPGTs for transform learning. Our experimental results demonstrated that SPGTs offer significant gains over KLT-based methods in data-limited scenarios, and joint clustering produces more efficient transforms than tree-structured clustering. Future work will focus on implementing and evaluating the proposed joint transform design approach within state-of-the-art video codecs.
\label{sec:foot}

% \vfill
% \pagebreak

\bibliographystyle{IEEEbib}
\bibliography{GSP,VideoCoding}

\end{document}

%% file: use_packages.tex
\usepackage{cite}
\usepackage{url}
\usepackage{textcomp}

\usepackage{multirow}
\usepackage{amsmath,amssymb,amsfonts}
\usepackage{stfloats}
\usepackage{nicefrac}

\usepackage{graphicx}
\usepackage[caption=false, font=footnotesize]{subfig}
\usepackage{xcolor}
\usepackage[acronym]{glossaries}

\usepackage{algorithm}
\usepackage{amsmath}
\usepackage{mathtools}

\usepackage{enumitem}
\usepackage{array}
\usepackage{amsthm}
\usepackage[titletoc,toc,title]{appendix}
\usepackage[bookmarksopen=true]{hyperref}
\usepackage[capitalise,nameinlink]{cleveref}

\usepackage{nccmath}
\usepackage{bookmark}

\usepackage{csquotes}
\usepackage{url}
\usepackage{mathrsfs}

\usepackage{algorithmic}
\usepackage{scrextend}

%% file: my_def_preample.tex
\graphicspath{Figures/}
\newcommand{\tr}{^\top}			% transpose of a matrix
\allowdisplaybreaks[4]					% allow breaks for equations
\long\def\comment#1{}

\newfont{\bbb}{msbm10 scaled 700}

\newcommand{\xv}{{\bf x}}
\newcommand{\yv}{{\bf y}}
\newcommand{\zv}{{\bf z}}

\newcommand{\Cm}{{\bf C}}

\newcommand{\Fm}{{\bf F}}
\newcommand{\Gm}{{\bf G}}

\newcommand{\Id}{{\bf I}}

\newcommand{\Lm}{{\bf L}}

\newcommand{\Pm}{{\bf P}}

\newcommand{\Rm}{{\bf R}}
\newcommand{\Sm}{{\bf S}}
\newcommand{\Tm}{{\bf T}}

\newcommand{\Xm}{{\bf X}}

\newcommand{\Cc}{{\cal C}}

\newcommand{\Gc}{{\cal G}}

\newcommand{\Sc}{{\cal S}}
\newcommand{\Tc}{{\cal T}}

\renewcommand{\vec}{{\rm vec}}

\theoremstyle{remark}